# A coarse-grained model with implicit salt for RNAs: predicting 3D structure, stability and salt effect

Ya-Zhou Shi[†], Feng-Hua Wang[†], Yuan-Yan Wu, and Zhi-Jie Tan[*]

Department of Physics and Key Laboratory of Artificial Micro- and Nano-structures of Ministry of Education, School of Physics and Technology, Wuhan University, Wuhan 430072, China

Short version of the title: A coarse-grained model for RNAs

## ABSTRACT

To bridge the gap between the sequences and 3-dimensional (3D) structures of RNAs, some computational models have been proposed for predicting RNA 3D structures. However, the existed models seldom consider the conditions departing from the room/body temperature and high salt (1M NaCl), and thus generally hardly predict the thermodynamics and salt effect. In this study, we propose a coarse-grained model with implicit salt for RNAs to predict 3D structures, stability and salt effect. Combined with Monte Carlo simulated annealing algorithm and a coarse-grained force field, the model folds 46 tested RNAs ($\leq$ 45 nt) including pseudoknots into their native-like structures from their sequences, with an overall mean RMSD of 3.5 Å and an overall minimum RMSD of 1.9 Å from the experimental structures. For 30 RNA hairpins, the present model also gives the reliable predictions for the stability and salt effect with the mean deviation ~ 1.0℃ of melting temperatures, as compared with the extensive experimental data. In addition, the model could provide the ensemble of possible 3D structures for a short RNA at a given temperature/salt condition.

---

† The authors contributed equally to the work.

* To whom correspondence should be addressed. Email: zjtan@whu.edu.cn

# I. INTRODUCTION

The central dogma of molecular biology stipulates that RNA plays a pervasive role in gene regulation and expression.[1] Within the past few years, discoveries of various noncoding RNAs have led to new insights in the importance of RNAs in many cell life processes. These diverse RNA molecules include ribozymes which catalyze cleavage or ligation of RNA backbone,[2] small interference RNAs which induce gene silencing[3,4] and riboswitches which control gene expression by directly sensing the levels of specific small-molecule metabolites[5].

To perform the biological functions, RNAs generally adopt compact native tertiary structures. Although the knowledge of the spatial structures and dynamics of RNAs is a fundamental prerequisite to completely understand their functions,[6-11] to obtain high-resolution RNA structures by experimental methods such as X-ray crystallography and nuclear magnetic resonance spectroscopy is very time-consuming and expensive compared to the determination of RNA sequences. Furthermore, due to the high flexibility and high charge density of RNA backbone, RNA structures are very sensitive to the solution environment such as temperature and salt.[6,12-20] Therefore, to build three-dimensional (3D) structures of RNAs remains an important challenge, especially at non-physiological solution conditions.[6,21,22] In an alternative way, computational modeling becomes very important to predict RNA 3D structures, and then to understand their biological functions.[23-34]

In recent years, some computational models have been developed for predicting RNA 3D structures[35-67]. The graphics-based approaches, such as MANIP[35], S2S/Assemble[36,37] and RNA2D3D[38] can be used to model small- to large-size structured RNAs from their secondary structures. Although the approaches are quick and intuitive, they are limited by the requirements for the users' interactive operation and expert knowledge. Based on the similarity between the structures of the evolutionarily related RNAs, another series of models have been developed such as ModeRNA[41] and RNABuilder[42]. These models can predict the structures for large RNAs while it may be difficult to find a proper template in databases for a particular RNA. In addition to the above knowledge-based models[35-44,51], there are some physics-based models[45-50,52,55-67], which simulate the folding process through sampling the conformations with minimizing free energy. The atomistic models such as MC-Fold/MC-Sym pipeline[45] and FARNA[46,47] can make promising predictions for 3D atomistic structures, while generally either depend on the knowledge of the secondary structure or only treat small RNAs. Since an atomistic structure model generally involves huge number of freedom degrees,[68] the simplified coarse-grained (CG) models[55-67,69-77] such as NAST[56], iFold[58,59] and Vfold[60,61] have been developed by treating a group of functional atoms as a single bead. The NAST[56] can be used to model large RNA molecules (>100 nt) based on the known secondary structures and certain tertiary contacts. The iFold[58,59] can predict 3D structures of small RNAs from sequences with the use of the discrete molecular dynamics sampling. The Vfold[60,61] can make reliable predictions on the free energy landscape of RNA pseudoknots. Although some of the CG models could take salt into account, it is still difficult for them to quantitatively predict the stability of various RNAs in salt solutions.

Despite the advances, there are few RNA 3D structure prediction models with the ability to quantitatively predict the thermodynamic properties of RNAs, especially in salt solutions. Recently, a CG model was developed to quantitatively predict the folding thermodynamics of RNA pseudoknots.[75,76] However, it could not give reliable prediction for RNA 3D structure from sequence. Therefore, it is imperative to propose a model to predict 3D structures of RNAs at a given salt/temperature condition.

Here, beyond the existed CG models, we propose a new CG model for short RNAs to predict 3D structures, stability and salt effect using the Monte Carlo (MC) simulated annealing algorithm. In the model, each nucleotide is represented by three beads representing phosphate, sugar and base, respectively. The knowledge-based potentials are implemented for the bonds, angles, and dihedrals for the CG beads, and the sequence-dependent base-pairing and base-stacking interactions, and electrostatic interaction are also included. The CG nature of the model, as well as the high-efficiency of MC algorithm, enables us to predict the 3D structures and stability of short RNAs at different salt conditions. The model works well for extensive short RNAs, and it could be extended to treat larger RNAs in solution containing $Mg^{2+}$ which was shown to play a critical role in RNA folding and functions.

## II. MODEL AND METHODS

### A. CG structural model

To reduce the complexity of RNA molecules, for each nucleotide, we adopt a reduced representation with three beads which correspond to the phosphate, sugar and base, respectively. As shown in Fig. 1a, the three beads are placed on the existing atoms: the phosphate (P) bead and the sugar (C) bead are placed at the P and C4' atom position, and the purine base bead and the pyrimidine bead are placed at the N9 and the N1 atom position, respectively.[61] P, C (C4') and N (N1 and N9) beads are treated as the spheres with van der Waals radii of 1.9 Å, 1.7 Å and 2.2 Å,

respectively.[68,78]

## B. Force field and optimization

### 1. Energy functions

The implicit-solvent/salt force field in our model is expressed as a summation of bonded and nonbonded terms by

$$U = U_b + U_a + U_d + U_{exc} + U_{qq} + U_{bp} + U_{bs}. \qquad (1)$$

The bonded term, namely the first three terms in Eq. 1, is a summation of potentials for virtual bond length $U_b$, bond angle $U_a$, and dihedral $U_d$, which were initially parameterized by the statistical analysis on the available 3D structures of 157 RNA molecules in the Protein Data Bank (PDB, http://www.rcsb.org/pdb/home/home.do) and their PDB codes are listed in Table SI[79]. The function forms of $U_b$, $U_a$, $U_d$, are given by Eqs. S3-S5 in supplementary material[79]. Since the structural features of RNAs are different for stems (helical) and single-strands/loops (nonhelical), two sets of parameters are calculated for the potentials of virtual bond length $U_b$, bond angle $U_a$ and dihedral $U_d$ for stems and single-strands/loops, named as Para$_{helical}$ and Para$_{nonhelical}$, respectively. As Figs. S1-S3 shown, the distributions of bond length/angle/dihedral for nonhelical parts are slightly broader than that of helical parts.[79] It is because that the bases in the loops of native structures in PDB are sometimes stacked with their neighbours, and consequently, the loops generally have some features of stems (helical parts). *The Para$_{helical}$ are used in the folding process to describe the folding of free RNA chains, and during the structure refinements on the initial folded structures, the*

*Para*$_{helical}$/*Para*$_{nonhelical}$ are used for helical/nonhelical regions, respectively.

The remaining terms of Eq. 1 describe various pairwise nonbonded interactions. The excluded volume energy $U_{exc}$ between the CG beads is modeled by a purely repulsive Lennard-Jones potential; see Eq. S8[79]. The electrostatic interaction $U_{qq}$ between phosphate groups is accounted for by the Debye-Hückel potential combined with the concept of counterion condensation:

$$U_{qq} = \sum_{i<j}^{N_P} \frac{(Qe)^2}{4\pi\varepsilon_0 \varepsilon(T) r_{ij}} e^{-\frac{r_{ij}}{l_D}}, \quad (2)$$

where $r_{ij}$ is the distance between two phosphate beads $i$ and $j$. $N_p$ is the number of phosphate beads. $l_D$ is the Debye length which is related to salt concentration and temperature; see Eq. S10[79]. $\varepsilon_0$ is the permittivity of free space and $\varepsilon(T)$ is an effective temperature-dependent dielectric constant[75,80]: $\varepsilon(T) = 87.74 - 0.4008T + 9.398 \times 10^{-4}T^2 - 1.41 \times 10^{-6}T^3$, where $T$ is in °C. $Qe$ in Eq. 2 is the net charge of each phosphate bead, where $e$ is the elementary charge and $Q$ is given by $Q=b/l_B$.[75,81] Here $b$ is the charge spacing on RNA backbone and $l_B$ is the Bjerrum length.

The base-pairing interaction between nucleotides is an important interaction in the folding of RNAs. For simplicity, we use three types of distances to model the orientation-dependent hydrogen-bonding interaction between specific nucleotide pairs (Fig. 1b), including the canonical Watson-Crick base pairs (G-C and A-U) and the wobble base pairs (G-U). For two nucleotides $i$ and $j$, if the distance $r_{N_iN_j}$ between two base beads $N_i$ and $N_j$ satisfies the paired criteria: $a_1<r_{N_iN_j}<a_2$, the hydrogen-bond formed and the corresponding base-paring potential is given by

$$U_{bp} = \sum_{i<j-3}^{N_{bp}} \frac{\varepsilon_{bp}}{1 + k_{NN}\left(r_{N_iN_j} - r_{NN}\right)^2 + k_{CN}\sum_{i(j)}\left(r_{C_iN_j} - r_{CN}\right)^2 + k_{PN}\sum_{i(j)}\left(r_{P_iN_j} - r_{PN}\right)^2}, \quad (3)$$

where $\varepsilon_{bp}$ (<0) is the interaction strength, which depends on the number of formed hydrogen bonds, and $\varepsilon_{AU} = \varepsilon_{GU} = \gamma\varepsilon_{GC}$ where $\gamma$ describes the ratio of pairing strength between different types of bases.[58,65,66,82] In base-pairing potential, the distance $r_{C_{i(j)}N_{j(i)}}/r_{P_{i(j)}N_{j(i)}}$ between sugar bead $C_{i(j)}$ or phosphate bead $P_{i(j)}$ and base bead $N_{j(i)}$ are used to restrict the orientation between the pairing nucleotides[58,82]; see Fig. 1b. The coefficients $k_{NN}$, $k_{CN}$ and $k_{PN}$ are the corresponding energy strength of the three base-pairing constraints and $r_{NN}$, $r_{CN}$ and $r_{PN}$ are their optimal distances obtained from the known structures; see Fig. S4a.[79] $\Sigma_{i(j)}$ in Eq. 3 stands for the summation over $i$ and $j$. In the model, one nucleotide cannot become paired with more than one nucleotide. Although such a strict constraint for base pairing requires a large $|\varepsilon_{bp}|$ to overcome the entropy change due to base pairing, it is an effective potential to efficiently take two complementary bases into preferable helix.

The base-stacking interaction provides a strong force in stabilizing RNA secondary structure. In the model, if two nearest neighbour bases $i$ and $i+1$ are paired with other neighbour bases $j$ and $j$-1 respectively, the base stacking is formed as shown in Fig. 1c. The base-stacking potential can be given by

$$U_{bs} = \frac{1}{2}\sum_{i,j}^{N_{st}} |G_{i,i+1,j-1,j}| \left\{ \left[ 5\left(\frac{\sigma_{st}}{r_{i,i+1}}\right)^{12} - 6\left(\frac{\sigma_{st}}{r_{i,i+1}}\right)^{10} \right] + \left[ 5\left(\frac{\sigma_{st}}{r_{j,j-1}}\right)^{12} - 6\left(\frac{\sigma_{st}}{r_{j,j-1}}\right)^{10} \right] \right\}, \quad (4)$$

where $\sigma_{st}$ is the optimal distance of two neighbour bases in the known helix structures shown in Fig. S4b.[79] $G_{i,i+1,j-1,j}$ are the strength of base stacking energy, which was derived from the combined analysis of available thermodynamic parameters and the MC algorithm

$$G_{i,i+1,j-1,j} = \Delta H - T(\Delta S - \Delta S_c). \quad (5)$$

Here $T$ is the absolute temperature in Kelvin. $\Delta H$ and $\Delta S$ are the RNA thermodynamic parameters associated with stacking between two adjacent base pairs and have been experimentally measured by Turner and colleagues.[83,84] Since a part of entropy change due to base-pair stacking is naturally included in the MC sampling process, this part of entropy change $\Delta S_c$ should be removed from $\Delta S$. To estimate $\Delta S_c$, we performed lots of MC simulations for one free base pair of an A-form double-stranded RNA at different locations and calculated the entropy change when it stacks with neighbour base pair; see more details on the calculation of $\Delta S_c$ in Eq. S15 and Fig. S5 in supplementary material.[79]

## 2. Determination of the parameters of energy functions

For the above described potentials, the initial parameters are directly obtained from the statistical analysis on the known structures (157 RNAs listed in Table SI).[79] Afterwards, the parameters are optimized through the comparisons with the experiments and the consequently slight adjustment.[85,86] In practice, five RNA hairpins (PDB code: 1q75, 1i3x, 1bn0, 2kd8, 28sr listed in Table SI and SIV)[79] are used for 3D structure comparisons and three other RNA hairpins (RH1, RH18, RH23 listed in Table I) are used for the comparisons on stability. The optimized parameters have been tabulated in Tables SII and SIII.[79] It is noted that the parameters for bond length, bond angle and dihedral do not differ significantly from the initial parameters. The base-pairing strength $\varepsilon_{bp}$=-3.5 kcal/mol (Table SIII) in Eq. 3 sounds a little large, while appears not so strong in RNA folding process due to the strict constraints of distance for base pairing, and the formation of helix is mainly determined by the $T$-dependent base stacking (see Eq. 4). The charge spacing $b$ in Eq. 2 on backbone is taken as 5.5 Å in the calculation, a slightly smaller value than the distances between two adjacent phosphate groups of single-stranded RNA (~6.0 Å) since the chain is generally bent rather than straight. *The optimized parameters from 5 hairpins for structure and 3 hairpins for stability are then used to make predictions on 3D structure for 46 RNAs and on stability for 30 RNA hairpins.*

### C. Simulation algorithm

With the above energy functions for the CG beads in RNAs, the present CG model can be employed to fold RNAs into native-like structures with the MC simulated annealing algorithm[87], which can effectively avoid the trap in local energy minima and has been used to predict the folding of proteins and RNAs[88].

The MC algorithm is performed as follows. Firstly, an initial random conformation is generated from RNA sequence at initial high temperature. Secondly, at each temperature, two different types of moves for the RNA chain are performed: subtle translation and the pivot move, which has been demonstrated to be rather efficient in sampling conformations of a polymer.[78,89] The change after each move is accepted or rejected according to the standard Metropolis algorithm.[78,89] Finally, after long enough steps for the system to reach equilibrium, the temperature is lowered based on the exponential scheme and the previous process is repeated until the target temperature is reached.[87]

## III. RESULTS AND DISCUSSION

Based on the implicit-solvent/salt force field for CG beads, we employ the model to predict 3D structures, stability and salt effect for various short RNAs. As compared with the experimental structures and the experimental thermodynamics data, the present model can make overall reliable predictions.

### A. RNA 3D structure prediction

The 3D structures of 46 RNAs with length ≤ 45 nt are predicted with the present model and compared with the experimentally measured structures. The RNA structures include hairpins, hairpins with bulge loop, hairpins with internal loop, and pseudoknots and the PDB codes of all these RNAs are listed in Table SIV. It should be noted that more than one half of the tested RNAs are not included in the dataset (Table SI) for obtaining the initial parameters of energy functions and only 5 RNAs (PDB code: 1q75, 1i3x, 1bn0, 2kd8, 28sr) are in the dataset for optimizing the parameters of energy functions. Here, the solution contains 1M NaCl and the RNAs are nearly fully neutralized by ions during the structure prediction process; see Eq. 2. In the following, we first select an RNA hairpin as an example to show how it folds in the present algorithm and afterwards, we will show our predictions for 46 RNAs including RNA hairpins with bulge, hairpins with internal loop and pseudoknots, and the comparisons with experimental structures. Finally, we will compare the present model with other two representative models: FARNA[46] and MC-Fold/MC-Sym pipeline[45].

1. *Folding process of a paradigm RNA hairpin*

*Initial 3D structure prediction from random chain.* To show the folding process in the present model, we select an RNA hairpin (PDB code: 1u2a; sequence: 5'-GGUCAGUGUAACAACUGACC-3') with a 6-nt loop and a 7 base-pair (bp) stem as a paradigm. First, a random chain (e.g., structure A in Fig. 2a) is generated from the sequence based on bonded potentials. Next, starting from this extended random conformation, the MC simulated annealing algorithm is performed from an initial high temperature to the target temperature (e.g., 298K). In the MC simulation annealing process, the energy of the hairpin reduces with the decrease of temperature and finally it fluctuates around a certain low value at room temperature, as shown in the top panel of Fig. 2a. Simultaneously, the hairpin folds into native-like structures (e.g., structure C in Fig. 2a) at the target temperature from the initial random chain step by step; see the middle panel and bottom panel of Fig. 2a. It is necessary to point out that only the $Para_{nonhelical}$ are employed for bonded potentials in the process to describe the folding of a free chain.

*Refinement on initial folded 3D structure.* After the annealing process, the initial native-like 3D structure of the hairpin is predicted. However, due to the $Para_{nonhelical}$ are insufficient to perfectly depict the more standard geometry of helical parts, the further structure refinement should be implemented. During the structure refinement, the $Para_{helical}$ (shown in Table SII)[79] are introduced to replace the parameters used before for the base-pairing regions (stems) in the initially folded structure. In addition, the RNA conformational changes are implemented by the subtle translation move of a single bead rather than the wide pivot moves used in the above simulated annealing process for RNA chains. As shown in Fig. 2b, the energy of the refined structures of the hairpin (PDB code: 1u2a) is about 5.0 kcal/mol lower than that of the structures before the refinement (Fig. 2b, top panel).

*Evaluation of the predicted 3D structures.* The predicted 3D structures are evaluated by their root-mean-square deviation (RMSD)[90] calculated over C beads from the corresponding atoms C4' in the native structures in PDB, and the RMSDs between the predicted structures and the experimental structures can be calculated by VMD software[91]. Since the present model generally predicts a series of native-like structures, in the following, we will use mean and minimum RMSDs to evaluate the reliability of the predictions. The former is the averaged RMSD over all the equilibrium structures and the latter is the RMSD for the equilibrium structure closest to the native one. For the hairpin shown here (PDB code: 1u2a), the mean RMSD and the minimum RMSD between predicted structures and its native structure are 2.6 Å and 1.5 Å, respectively; see the middle panel and bottom panel of Fig. 2b.

2. *3D structures of 46 RNAs*

According to the above process, we predict the 3D structures for the 46 tested short RNAs listed in Table SIV,[79] where mean RMSD and minimum RMSD of each RNA are also listed. All the 46 RNAs fold into their near-native structures with the overall mean RMSD of 3.5 Å and the overall minimum RMSD of 1.9 Å from their native structures; see Table SIV.

*RNA hairpins.* Since an RNA hairpin is a simplest secondary structure of RNAs, we predicted 3D structures for 18 RNA hairpins with different lengths with the overall mean RMSD of 2.8 Å from the corresponding experimental structures. As Fig. 3a shown, the present model is very reliable for predicting the structures of stems, while the predicted loops are slightly deviated from the experimental ones.

*Hairpins with bulge loop.* RNA bulge loops, which interrupt one strand of a continuous double helix, occur frequently in the secondary structures of large RNAs generally as recognition sites. The overall mean RMSD between the predicted structures of the 9 hairpins with bulge loop and the experimental structures is 3.3 Å. The value of mean RMSDs for the hairpins with bulge loop is slightly larger than that for the hairpins without bulge loop possibly because that bulge loops usually distort the RNA backbone and cause higher flexibility of RNAs; see Fig. 3b.

*Hairpins with internal loop.* An internal loop, which separates an RNA into two helical regions, generally causes strong distortion and high flexibility of the RNA. For the 15 tested hairpins with internal loop, the overall mean RMSD is 4.0 Å, a larger value than those of hairpins with and without bulge loop; see Table SIV.[79] Obviously, if a helix region is modeled to be roughly in the right place, while another helix region is angled relatively to the correct orientation, the conformation could propagate to produce large RMSD values even with modest degrees of angular deflection (Fig. 3c). Beyond that, as important functional fragments (e.g., as binding sites), internal loops usually have specific structures, which may contain noncanonical motifs/base-pairs.[9]

*Pseudoknots.* An RNA pseudoknot is defined as the structure with the base pairing between a loop and other single-stranded regions of an RNA. Pseudoknots play diverse fundamental roles in the control of viral replication, in structural organization of complex RNAs and in the self-cleaving ribozyme catalysis.[92,93] Four pseudoknots have been tested by the present model and the mean RMSDs for these pseudoknots are 4.2 Å, 5.2 Å, 4.2 Å, and 5.4 Å, respectively; see Table SIV.[79] Although the model accurately predicts secondary structures of these pseudoknots, it only moderately captures their 3D structures with the slight deviation from experimental structures for pseudoknot loops; shown in Fig. 3d. A possible reason is that the present model ignores some interactions in RNA pseudoknots, such as the single-stranded base stacking and some specific hydrogen bonds between bases and backbone.[93-95] In addition, the model predicted a two-step folding for the pseudoknots: a chain would firstly fold into an intermediate state of hairpin and then to the final state of pseudoknot, which is in accordance with the previous studies[96,97].

Fig. 3 shows the predicted 3D structures (ball-stick) with the mean and minimum RMSDs and the experimental structures for four typical RNAs. The overall comparison of the predicted structures (ball-stick in Fig. 3) with the experimental structures (cartoon in Fig. 3) shows that the present model is very effective to capture the 3D shapes of short RNAs. As shown in Fig. 4, the prediction accuracy of 3D structures by the present model decreases with the increase of RNA length (Fig. 4a), especially with the number of unpaired nucleotides (Fig. 4b) possibly due to their highly flexible nature. This suggests that the further improvement on the prediction accuracy requires the more accurate treatment on loops.

3. *Comparisons with previous models*

We test our predictions against two well-established 3D structure prediction models: FARNA[46] and MC-Fold/MC-Sym pipeline[45]. The RMSDs of our predicted structures are calculated over C beads from the corresponding C4' atoms in the native structures and the predicted structures are not further refined by all-atomistic molecular dynamics. Firstly, we make comparisons with the predictions from FARNA, and the RMSDs calculated over C4' atoms from FARNA are taken from Ref. 46. As shown in Fig. 5a, the average prediction accuracy (mean RMSD = 3.82 Å; minimum

RMSD = 2.37 Å) by the present model is not worse than that of FARNA (mean RMSD = 3.92 Å; minimum RMSD = 2.2 Å). Secondly, we further make comparisons between the present model and MC-Fold/MC-Sym pipeline. For all the above tested RNAs, we used the MC-Sym online server (option: model_limit =1000 or time_limit = 12h) to predict the best 3D structures with lowest score using the best 2D structures predicted by the MC-Fold (option: consider H-type pseudoknots and return the best 100 structures) and then we calculated the RMSDs of the predicted structures over C4' atoms from the corresponding atoms in their native structures. Fig. 5b and Table SIV[79] show that for the 46 RNAs (≤ 45 nt), the overall mean RMSD of structures from the present model is 3.5 Å, which is slightly smaller than 3.9 Å, the mean RMSD of the top 1 structures from the MC-Fold/MC-Sym pipeline.

Beyond 3D structure prediction, the present model could also make reliable predictions on the stability of RNA hairpins and the salt effect.

## B. Stability of RNA hairpins

The folding thermodynamics of RNAs are important for unravelling structure-function relationships for RNAs. To obtain accurate thermodynamic parameters of RNAs, there are many thermal denaturation experiments and theoretical modeling for RNA hairpins[98-106] or duplexes[80,83]. In order to make further validation on the present model, we have performed the extensive simulations for 30 RNA hairpins to predict their melting temperatures $T_m$. Here the ionic condition is also assumed at 1 M NaCl where the RNAs can nearly get full-neutralization; see Eq. 2. All the sequences of 30 RNA hairpins tested here are listed in Table I.

We use the statistical average of the number of base pairs $P_{bp}(t,T)$ to characterize the state of hairpins and $P_{bp}(t, T)$ can be given by

$$P_{bp}(t,T) = \frac{1}{t}\sum_{t} N_{bp}(t,T), \tag{6}$$

where $t$ is the MC step. $T$ is temperature, and $N_{bp}(t, T)$ is the total number of base pairs at step $t$ and temperature $T$. Based on the equilibrium value of the number of base pairs at each temperature (e.g., transverse lines in Fig. 6a), the fraction of denatured base pairs $f(T)$ can be calculated and fitted to two-state model to obtain the melting temperature $T_m$

$$f(T) = 1 - \frac{1}{1 + e^{(T-T_m)/dT}}, \tag{7}$$

where $dT$ is an adjustable parameter.[83,101]

Fig. 6a shows that the number of base pairs changes at different temperatures in the simulation of the hairpin RH24 with a 10-nt loop and a 6-bp stem; see Table I. When the temperature is very high (~120℃), almost all of the base pairs are denatured (bottom panel) and when the temperature is low (~40℃), the hairpin is in a folded state (top panel). Around the melting temperature (~80℃), the folded and unfolded states appear alternately with approximately equal probability (middle panel), which illustrates the bistability in terms of base pair number near $T_m$. In addition, all RNA hairpins tested in this work (Table I) have the similar bistability (data not shown). Fig. 6b shows that the predicted data (symbols) and fitted melting curves (lines) for the hairpin RH24 mentioned above, and the melting temperature $T_m$ can be easily estimated from the curve. For RH24, the predicted $T_m$ (82.5℃) is very close to the experimental data (82.4℃); see Table I. In addition, an advantage of the present model is that it can provide a series of the 3D structures of RNA molecules at different temperatures $T$: native structures at low $T$, disordered chains at high $T$, and occasionally partially denatured structures at $T \sim T_m$; see Fig. 6b.

To examine the sequence effect, 30 RNA hairpins with extensive sequences have been studied with the present model. The sequences with the corresponding melting temperatures are listed in Table I, which shows that our predictions on $T_m$ agree well with the experimental data with the mean deviation of 1.1°C over the extensive sequences. In addition, Fig. 6c shows the melting curves for the hairpins of RH6, RH18 and RH23, and the agreement between our predictions and available experimental data[98,103] suggests that the present model can reliably predict the denature processes of RNA hairpins. However, for unusually stable RNA hairpins such as hairpins with GA mismatches and tetraloops, the present model cannot give accurate predictions on stability. This may be because that the special hydrogen-bond and base stacking interactions in loops are not accounted for in the present model.

## C. Salt effect in RNA hairpin stability

Since RNAs are strongly charged polyanionic chains, there is strong intrachain Coulombic repulsion during RNA folding process. The counterions in solutions are critical to RNA folding because the ions can neutralize backbone charges and consequently favour the folding.[12-14,107,108] Although the present model can conveniently involve the explicit salt ions, for simplicity and efficiency, we combine the DH theory with the concept of ion binding of Manning[75,81] by a DH potential between the reduced backbone charges; see Eq. 2. In the following, firstly, we will predict the structures of HIV-1 TAR RNA at different $Na^+$ concentrations ($[Na^+]$'s) and make the comparison with the available experiments[109]. Afterwards, we will study the $[Na^+]$-dependent stability in comparisons with the available experimental data[102-106].

*1. Na$^+$-dependent conformational change of HIV-1 TAR*

The transactivation response element (TAR) RNA from the human immunodeficiency type I virus (HIV-1) is a hairpin (29-nt) with a 3-nt bulge loop and its 3D structures are strongly dependent on counterions.[109] To examine the Na$^+$-dependent RNA structure change, Casiano-Negroni et al. have experimentally derived the Na$^+$-dependent angle between two stems of a tetraloop HIV-1 TAR variant.[109] To directly compare with the experimental data, we predict the 3D structures of the TAR variant at different [Na$^+$]'s and calculate the angles between two stems separated by the bulge. For each predicted structure, the two stems can be approximately fitted to canonical A-form helices and the central axises of the helices can be derived with the use of the Program Curves+[110]. Based on these central axises, the inter-helical angles for all predicted TAR variant structures at different [Na$^+$]'s can be calculated. As Fig. 7a shown, the angles between two stems predicted by the present model agree well with the experiment data[109], especially at low [Na$^+$]'s. Furthermore, the model also gives the possible 3D structures of the TAR variant at different [Na$^+$]'s; see Fig. 7a.

*2. Na$^+$-dependent stability of RNA hairpins*

Here we employed the present model to examine the [Na$^+$] effect on the stability of six RNA hairpins RH23, RH24, RH25, RH27, RH29 and RH30 whose sequences are shown in Table I.[102-106] For each RNA hairpin, we perform the simulations at different temperatures over a broad range of [Na$^+$]'s. Based on the data from the simulations, we obtain the melting curves at different [Na$^+$]'s by fitting the calculated data to the two-state model; see Fig. S6.[79]

As shown in Figs. 7b, c and d, the increase of [Na$^+$] enhances the RNA hairpin folding stability, and the predicted melting temperatures agree well with experimental data[102-106] with the mean deviation of 0.9℃. Thus, the present model can give the quantitative predictions on the melting temperatures of RNA hairpins over a broad range of [Na$^+$]'s. Furthermore, the present model can provide the ensemble of probable 3D structures at different [Na$^+$]'s, as shown in Fig. 7b. Generally, the RNA duplexes adjacent to the terminal base pair are not stable, because the terminal base pair of a duplex could only stack with one base pair rather than stack with two nearest neighbour ones and can fluctuate with the less spatial constraints. This property of the duplex was captured in our model, in accordance with the previous experiments[83,84].

3. *Hairpin denaturing induced by high temperature and low salt*

Folding/unfolding of an RNA is closely related to the solution environments such as temperature and salt concentration as well as the forces acting on it.[76,111,112] Although increasing temperature or decreasing salt concentration both can trigger unfolding of RNAs, the mechanisms are not the same.[112] For the hairpin RH24 (Table I), we calculated the statistical distributions of the end-to-end distances at different temperatures ([Na$^+$] = 1M; Fig. 8a) and different [Na$^+$]'s (T = 70℃; Fig. 8b). As Fig. 8b shown, although the melting transitions induced by high T and low [Na$^+$] both exhibit the two-state transition, the denatured states at low salt are different from those at high temperature; see the inset figures in Figs. 8a and 8b. The distributions of end-to-end distance show that the denatured structures at low salt become more extended with the decrease of [Na$^+$], while those at high temperatures appear independent on temperature. Such difference comes from the different mechanisms of RNA denatured induced by high T and low salt. The transition induced by low salt is mainly caused by the intrachain electrostatic repulsion, while that induced by high-T mainly results from the conformational entropy of RNA chain.

4. *Free energy barrier at $T_m$ versus [Na$^+$]*

In order to illustrate the salt effect on the denatured transition of RNA hairpins, we calculated the normalized populations ($P(N_{bp})$) of base pair number ($N_{bp}$) at $T_m$ and different [Na$^+$]'s for RH24; see Fig 8c. The relative free energy at $T_m$ can be calculated by $\Delta G = -k_B T \ln[P(N_{bp})]$; see Fig. 8d for three different [Na$^+$]'s. As shown in Fig. 8d, the largest free energy barrier between folded and unfolded states is at the formation of first base pair, and the relative free energy decreases as more base pairs are formed. This is reasonable since the formation of first base pair generally leads to a great loss of chain conformational entropy. Once the first base pair is formed, it will become easier for more base pairs to be formed.[10,11] Fig. 8d also shows that the free energy barrier between folded and unfolded states decreases as [Na$^+$] increases. Specifically, the free energy barrier is lowered ~1 kcal/mol when [Na$^+$] increases from 20mM to ~1M, which indicates that the transition is much easier at high salt. Such phenomenon mainly comes from the weaker intrachain electrostatic repulsion at higher [Na$^+$].

In the present work, we only studied the effect of monovalent salt. Multivalent counterions such as $Mg^{2+}$ stabilize RNA tertiary structure more effectively than monovalent ions,[12-15,113] which is beyond the description of the DH theory used in the present model due to the strong ion-ion correlations. To predict the stability of RNAs in multivalent ion solutions, explicit ions may need to be accounted for in the model and accordingly would increase the computation cost.[78]

## IV. CONCLUSIONS

In summary, we have developed and employed a new CG model with implicit salt for RNA folding with MC simulated annealing algorithm. The model enables us to predict 3D structures and stability of short RNA hairpins over a broad range of [$Na^+$]'s. The following are the major conclusions:

(1) The present model can predict the native-like 3D structures for RNA hairpins, with and without bulge/internal loops, and pseudoknots (≤ 45 nt), with an overall mean RMSD of 3.5 Å and an overall minimum RMSD of 1.9 Å from experimental structures. The prediction accuracy of the present model reduces with the increase of length of unpaired regions in RNAs.

(2) The present model can make reliable predictions on the stability of RNA hairpins such as melting temperature $T_m$ with the mean deviation of 1.1℃ from experimental data. Meanwhile, it can provide the ensemble of probable 3D structures of RNA hairpins at different temperatures.

(3) The present model can also predict the stability for RNA hairpins with the mean deviation of 0.9℃

over wide [Na$^+$]'s as compared with extensive experimental data, and simultaneously provide the ensemble of probable 3D structures at different [Na$^+$]'s.

Although our model makes the overall reliable predictions for native-like structures and stability of short RNA molecules over a broad range of [Na$^+$]'s, further improvements need to be made on the model to improve the predictive accuracy and to treat larger RNAs with complex structures. Firstly, the loop/unpaired regions of structures predicted by the present model are slightly deviated from experiment ones. A possible reason may be the neglect of some important intrachain interactions in the present model such as the self-stacking in single-stand chain, special hydrogen bonds involving phosphates and sugars, and mismatched base pairs.[114,115] Secondly, the model at the present version cannot effectively make prediction for large RNA molecules, especially for those with complex tertiary interactions, since the model doesn't take into account the non-cannonical base pairs and tertiary hydrogen bonds which are often present in large RNA structures.[114,115] Further improvement on the issue should contain more imperative interactions between atoms as well as enhance the sampling efficiency in conformational space[50] by involving the specific tertiary contacts[42,65,114-121]. Thirdly, the DH approximation used in the present model ignores the effects of ion-ion correlation and ion-binding fluctuation which can be important for multivalent ions (e.g., Mg$^{2+}$).[12-15,80,122] The model can be further extended to involve explicit ions while the conformation searching cost would increase accordingly.[78]

In addition, although the predicted CG structures contain the major topological information of RNA molecules, they are limited for practical applications due to the lack of atomistic details. It is necessary to reconstruct the all-atomistic structures based on the CG structures.[59,61,64,123]

Nevertheless, the present model could be a basis for a possible model for predicting 3D structures, the stability and salt effect for RNAs with complex structures.

## ACKNOWLEDGEMENTS

We are grateful to Shi-Jie Chen, Wenbing Zhang, and Song Cao for valuable discussions. This work was supported by the National Science Foundation of China grants (11074191, 11175132 and 11374234), the National Key Scientific Program (973)-Nanoscience and Nanotechnology (No. 2011CB933600), the Program for New Century Excellent Talents (NCET 08-0408) and the Fundamental Research Funds for the Central Universities (1103007). One of us (Y. Y. Wu) also thanks financial supports from the interdisciplinary and postgraduate programs under the Fundamental Research Funds for the Central Universities (201120202020007).

**FIGURES AND TABLES**

**FIGURE 1.** (a) Our coarse-grained representation for one fragment of RNA superposed on an all-atom representation. Namely, three beads are located at the atoms of phosphate (P, orange), C4' (C, green), and N1 for pyrimidine or N9 for purine (N, blue), respectively. The structure is shown with the PyMol (http://www.pymol.org). (b) The schematic representation for base-pairing, which is restricted by three types of distances: $N_i N_j$ (red), $C_{i(j)} N_{j(i)}$ (green) and $P_{i(j)} N_{j(i)}$ (orange); see Eq. 3. Here, we use the distance constraints rather than the angular constraints for convenient programing of the model. (c) The schematic representation for base-stacking. Dash-dotted line (blue): base-stacking between two adjacent base pairs.

**FIGURE 2.** The illustration of the present algorithm for the folding process (a) and structure refinement (b). (a) The time-evolution of the energy (top panel), the number of base pairs (middle panel) and the 3D structures (bottom panel) during the MC simulated annealing simulation of the RNA hairpin (PDB code: 1u2a). (b) The structural refinement of the predicted structure of the hairpin (PDB code: 1u2a) illustrated by the energy of optimized structures (top panel), the RMSDs between optimized structures and the native structure in PDB (middle panel), and the predicted 3D structures (ball-stick; bottom panel) of the hairpin with the minimum RMSD (1.5 Å) and the mean RMSD (2.6 Å) from the native structures (cartoon). The structures are shown with the PyMol (http://www.pymol.org).

**FIGURE 3.** The predicted 3D structures (ball-stick) with the minimum RMSD and the mean RMSD

for a hairpin (a), a hairpin with bulge loop (b), a hairpin with internal loop (c) and an RNA pseudoknot (d). The mean (minimum) RMSDs between the predicted structures and their native structures (cartoon) are 2.4 Å (1.1 Å), 3.4 Å (1.8 Å), 4.0 Å (2.4 Å) and 4.2 Å (2.7 Å), respectively. The RMSDs are calculated over C beads, and the structures are shown with the PyMol (http://www.pymol.org).

**FIGURE 4**. The scatter plots of mean (square) and minimum (circle) RMSDs between the predicted structures and the native structures as functions of RNA size (-nt) (a) and of the number (-nt) of unpaired nucleotides (b) for 46 tested RNAs for 3D structure prediction.

**FIGURE 5.** (a) Comparison of the RMSDs between the present model and the FARNA[46]. The RMSDs of structures including hairpins and pseudoknots predicted by FARNA are calculated over the C4' atom in the backbone and the data are taken from Ref. 46. (b) Comparison of the RMSDs between the present model and the MC-Fold/MC-Sym pipeline[45]. For each of the 46 tested RNAs, we use the MC-Fold/MC-Sym pipeline online tool (http://www.major.iric.ca/MC-Fold/) to test the accuracy of MC-Fold/MC-Sym and calculate the RMSD for the top 1 predicted structure over C4' atom in the backbone. The RMSDs in (a) and (b) of structures predicted by the present model are calculated over C beads from the corresponding C4' atoms in native structures.

**FIGURE 6.** (a) The time-evolution of the number of base pairs of the hairpin RH24 (shown in Table I) at different temperatures: 40℃ (top panel), 82.5℃ (middle panel) and 120℃ (bottom panel). The transverse lines are the average values of the number of base pairs. (b) The fraction of denatured

base pairs of the hairpin RH24 as a function of temperature. Symbols: the predicted data; Line: fitted to the predicted data through Eq. 7; Ball-stick: 3D structures at different temperatures shown with the PyMol (http://www.pymol.org). (c) The fractions of denatured base pairs for three RNA hairpins (RH6, RH18, and RH23 shown in Table I) as functions of temperature. Symbols: predicted data; Bold lines: fitted to the predicted data through Eq. 7; Dashed lines: experimental curves[98].

**FIGURE 7.** (a) The experimental (Ref. 109) and predicted inter-helical bend angle as functions of [$Na^+$] for tetraloop HIV-1 TAR variant at 25℃ and the corresponding typical 3D structures predicted by the present model. (b) The experimental (calculated from Table II in Ref. 103) and predicted fraction of denatured base pairs as functions of [$Na^+$] for RNA hairpin RH24 (in Table I) at 70℃ and the corresponding typical 3D structures. The 3D structures in (a) and (b) are shown with the PyMol (http://www.pymol.org). (c) and (d) The melting temperature $T_m$ as functions of [$Na^+$] for six RNA hairpins (shown in Table I). *Symbols*: experimental data (c) ■ RH23 (Ref. 103), ● RH24 (Ref. 103), ▲ RH30 (Ref. 105) and (d) ■ RH25 (Ref. 102), ● RH27 (Ref. 102), ▲ RH29 (Ref. 102). *Lines*: predictions for the corresponding RNAs.

**FIGURE 8.** Analysis on the melting of the hairpin RH24 whose sequence is shown in Table I. (a) The distributions of end-to-end distance for conformations at different temperature and 1M [$Na^+$]. (b) The distributions of end-to-end distance for conformations at different [$Na^+$]'s and 70℃. The inset in (a) and (b) show the zoomed portion of the figure in the interval of [20, 120]. (c) and (d) The normalized populations (c) and the free energy barriers (d) as functions of the number of base pairs at ~$T_m$ (82.4℃) and three different [$Na^+$]'s.

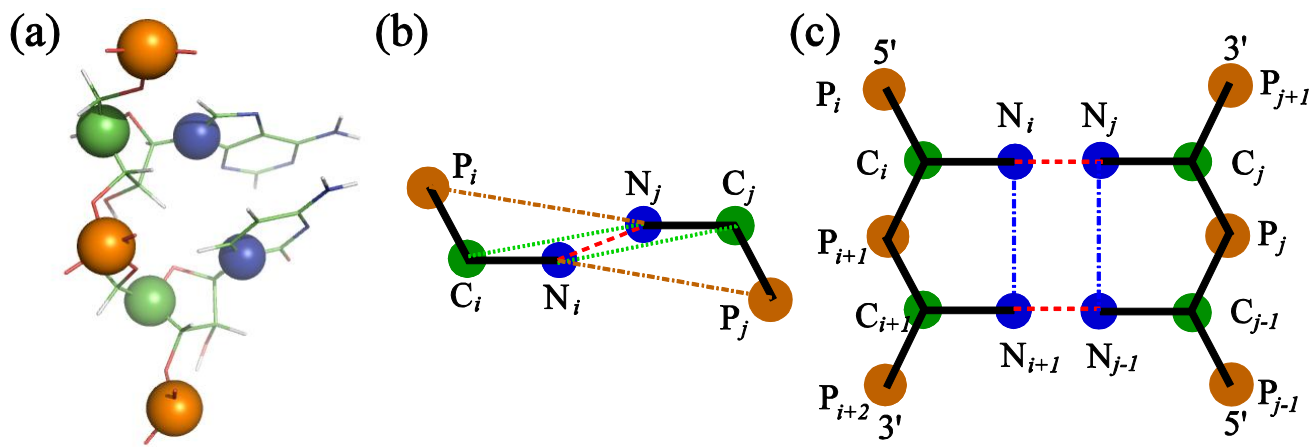

Figure 1:

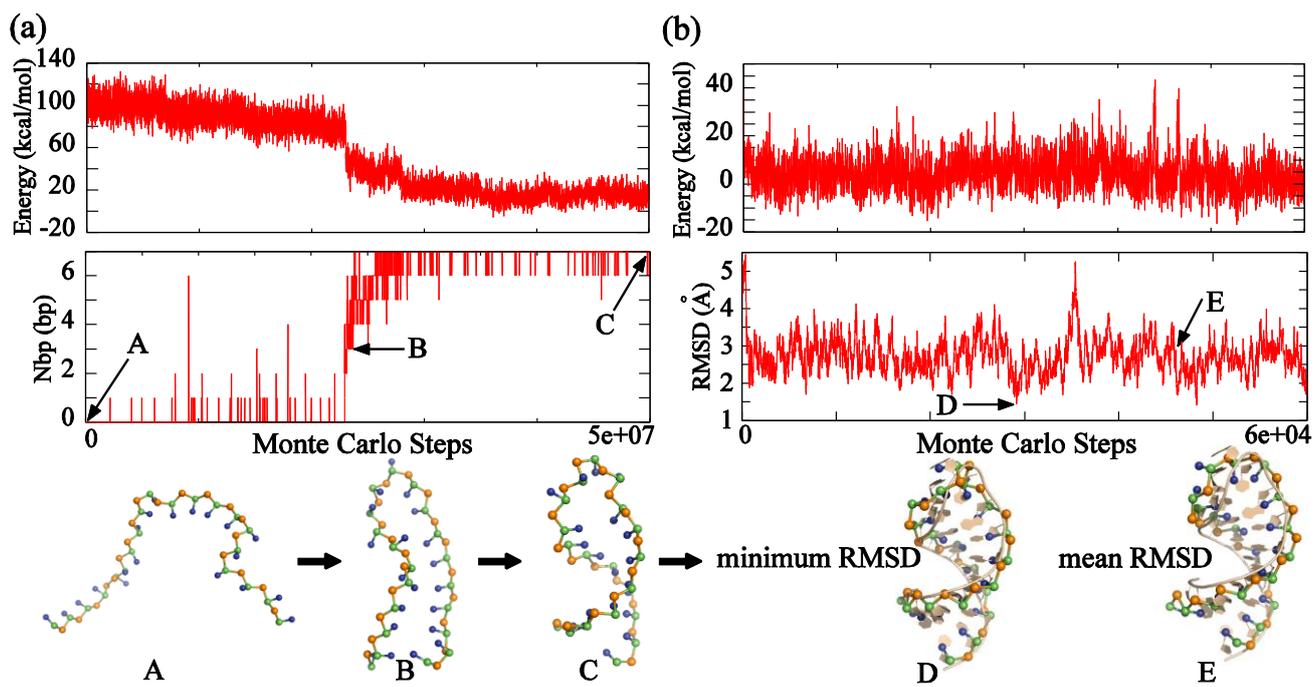

Figure 2:

(a) Hairpin (PDB code: 2kd8)
minimum RMSD      mean RMSD

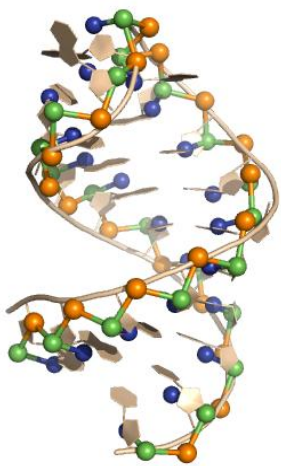 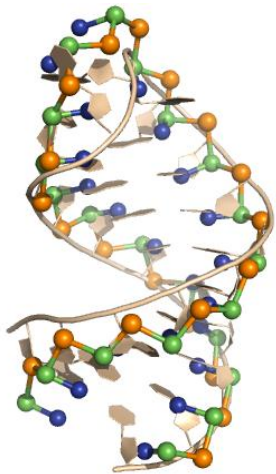

(b) Hairpin with bulge loop (PDB code:1zc5)
minimum RMSD      mean RMSD

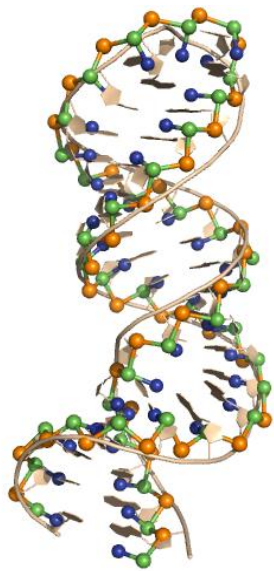 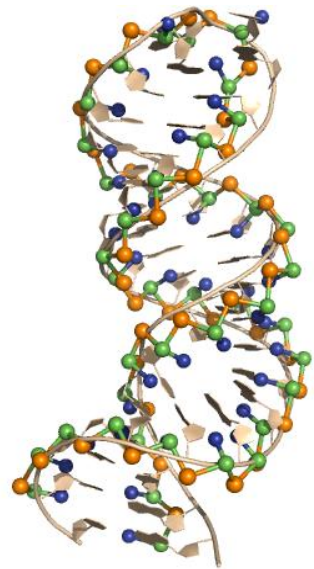

(c) Hairpin with internal loop (PDB code: 28sr)
minimum RMSD      mean RMSD

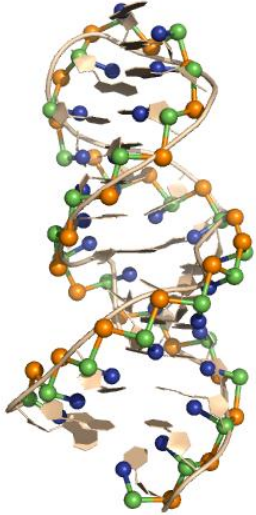 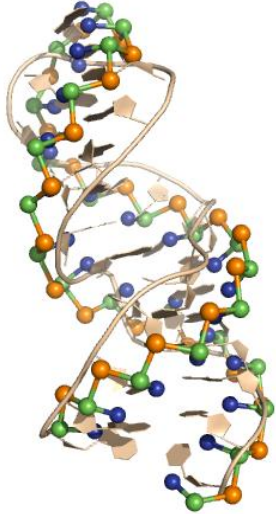

(d) Pseudoknot (PDB code: 2a43)
minimum RMSD      mean RMSD

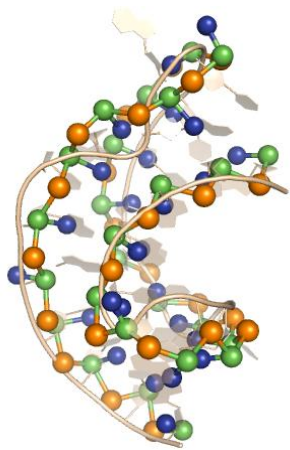 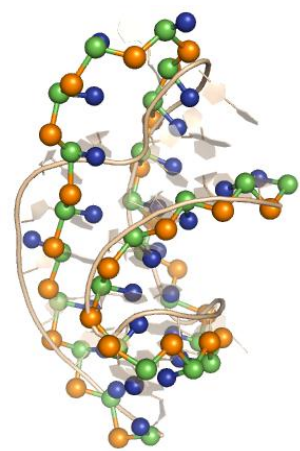

Figure 3:

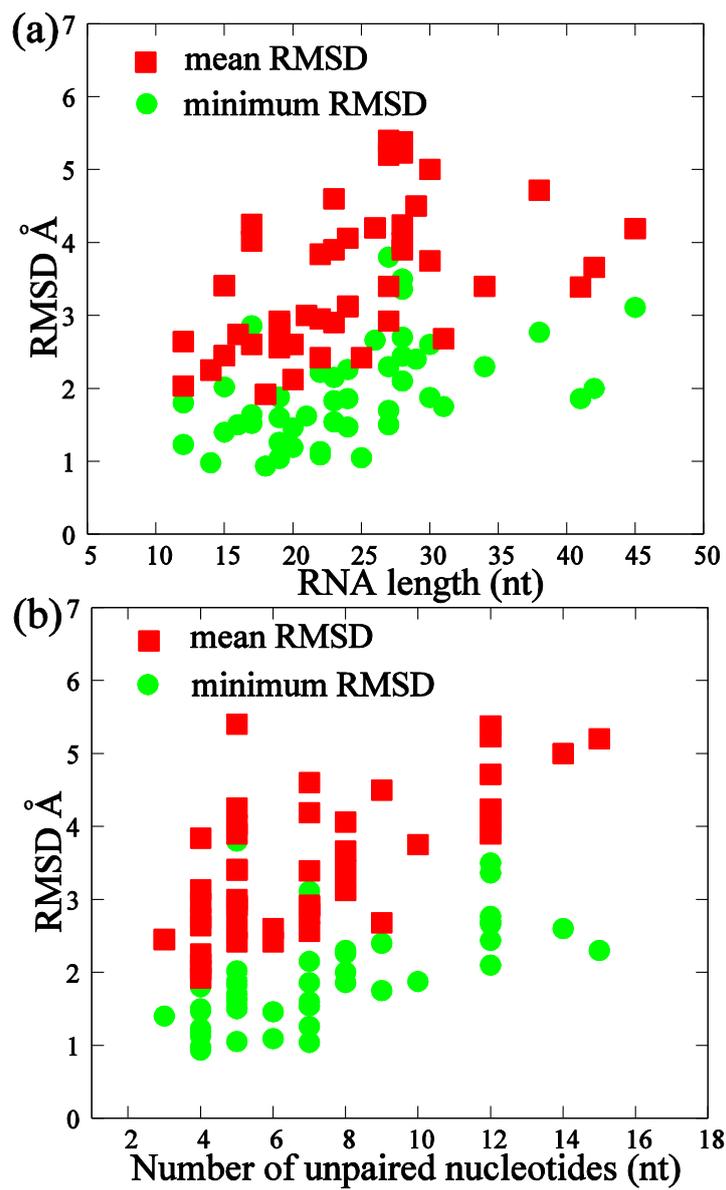

Figure 4:

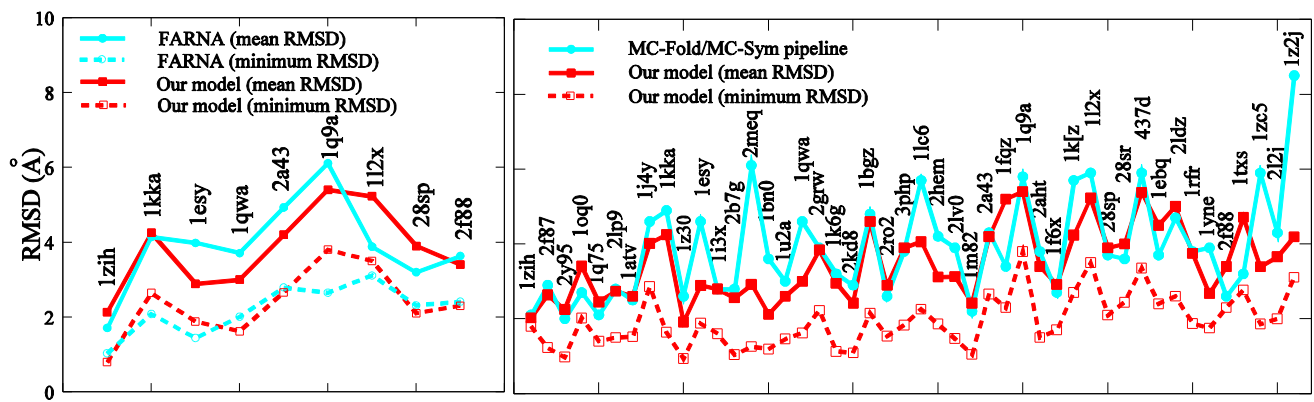

Figure 5:

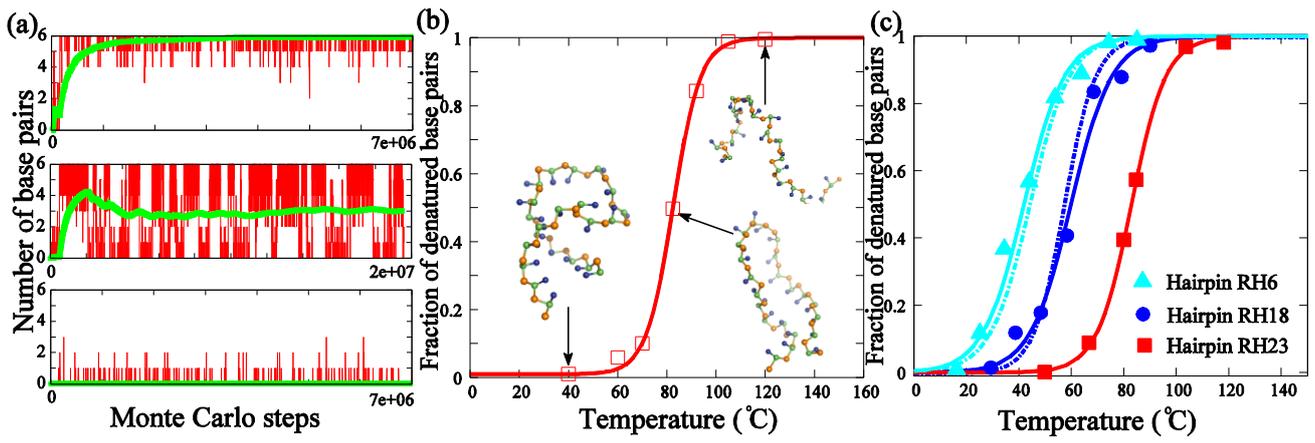

Figure 6:

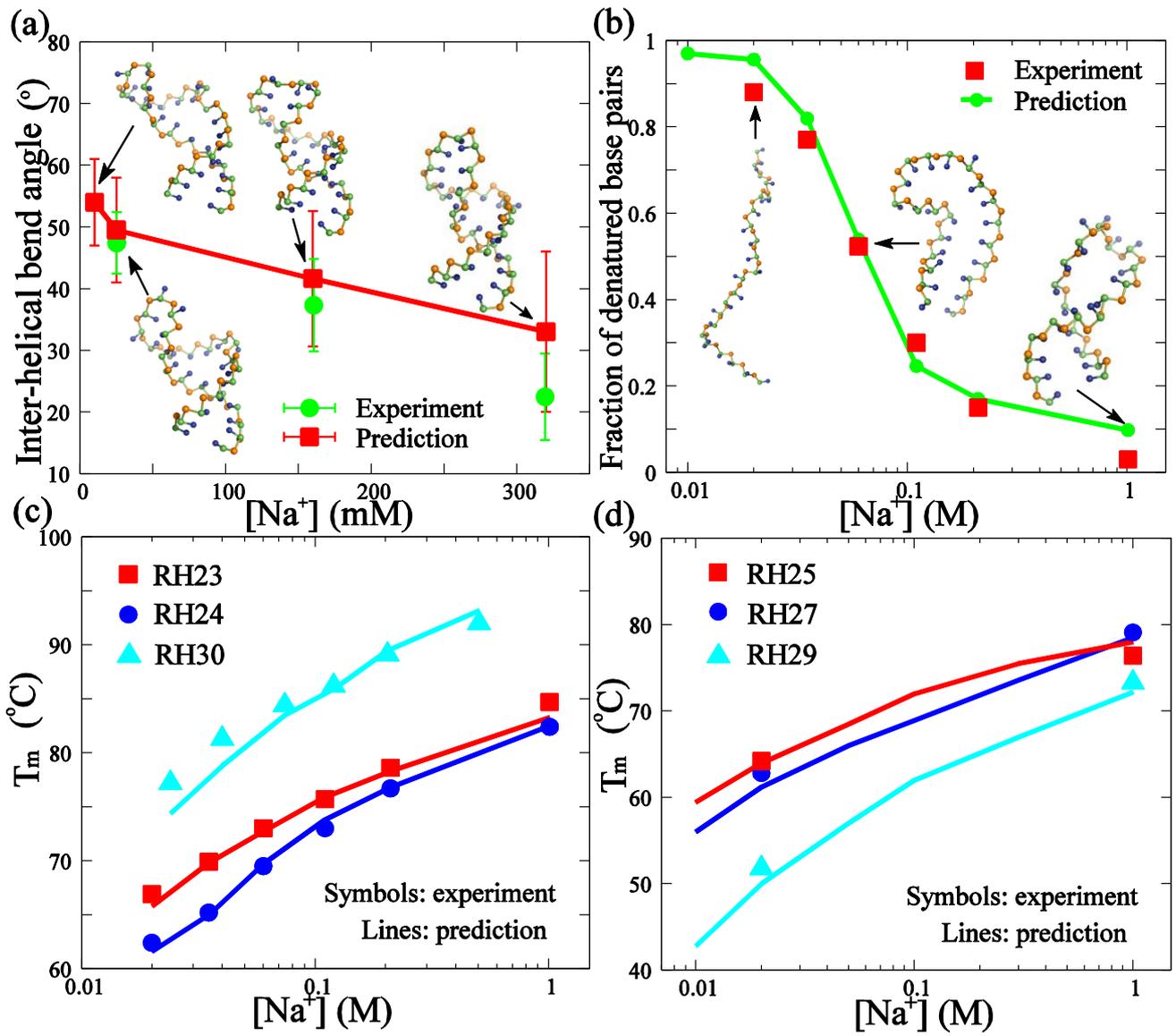

Figure 7:

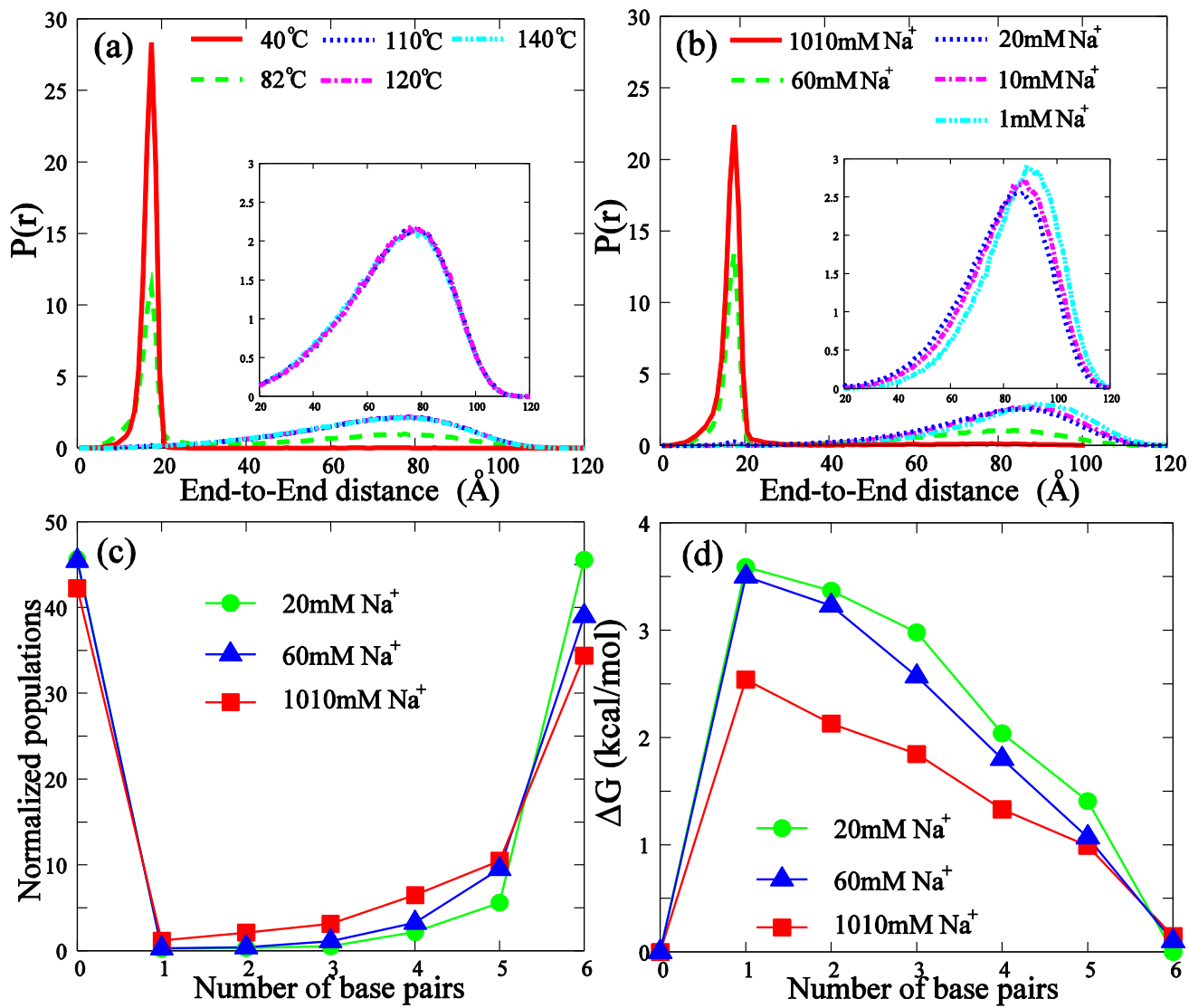

Figure 8:

**Table I.** The melting temperatures $T_m$ of 30 RNA hairpins at 1 M NaCl.

| RNA No. | Refs. | Sequences [a] (5' to 3') | Expt. (°C) | Pred. (°C) | Deviation (°C) |
|---|---|---|---|---|---|
| RH1 | 98 | GGC<u>AUAAUA</u>GCC | 64.2 | 62.2 | 2.0 |
| RH2 | 99 | GGC<u>AUAAAUA</u>GCC | 59.2 | 58.8 | 0.4 |
| RH3 | 98 | GGU<u>AUAAUA</u>ACC | 40.7 | 42.4 | 1.7 |
| RH4 | 101 | GGU<u>CUAAUC</u>ACC | 40.5 | 41.8 | 1.3 |
| RH5 | 101 | GGU<u>CUAAUA</u>ACC | 40.2 | 41.4 | 1.2 |
| RH6 | 98 | GGA<u>AUAAUA</u>UCC | 44.2 | 41.8 | 2.4 |
| RH7 | 98 | GGG<u>AUAAUA</u>CCC | 58.7 | 58.9 | 0.2 |
| RH8 | 99 | GGG<u>AUAAAUA</u>CCC | 57.3 | 57.0 | 0.3 |
| RH9 | 98 | GGU<u>AUAAUA</u>GCC | 42.8 | 41.8 | 1.0 |
| RH10 | 101 | GCG<u>AUAAUA</u>CGC | 56.1 | 55.0 | 1.1 |
| RH11 | 101 | GCG<u>UUAAUC</u>CGC | 52.7 | 54.2 | 1.5 |
| RH12 | 101 | GCG<u>CUAAUU</u>CGC | 50.8 | 52.5 | 1.7 |
| RH13 | 101 | GGU<u>CUAAUU</u>GCC | 40.7 | 42.0 | 1.3 |
| RH14 | 101 | GGU<u>AUAAUG</u>GCC | 42.9 | 42.5 | 0.4 |
| RH15 | 100 | GCA<u>CUAAUUU</u>GC | 38.9 | 39.2 | 0.3 |
| RH16 | 100 | GCA<u>AUAAUA</u>UGC | 42.6 | 42.8 | 0.2 |
| RH17 | 100 | GCA<u>UUAAUC</u>UGC | 42.9 | 40.5 | 2.4 |
| RH18 | 98 | AGGA<u>AUAAUA</u>UCCU | 57.0 | 58.8 | 1.8 |
| RH19 | 98 | AGGU<u>AUAAUA</u>GCCU | 56.6 | 58.0 | 1.4 |
| RH20 | 101 | GACG<u>UUAAUUU</u>GUC | 46.3 | 45.5 | 0.8 |
| RH21 | 101 | GACG<u>CUAAUUU</u>GUC | 46.4 | 46.6 | 0.2 |
| RH22 | 101 | GACG<u>CUAAUC</u>UGUC | 45.2 | 46.0 | 0.8 |
| RH23 | 103 | GAAGCC<u>AUUGCCCC</u>GGCUUC | 84.7 | 83.3 | 1.4 |
| RH24 | 103 | GAAGCC<u>AUUGCACUCC</u>GGCUUC | 82.4 | 82.5 | 0.1 |
| RH25 | 102 | GGGAUAC<u>AAA</u>GUAUCCA | 76.4 | 78.0 | 1.6 |
| RH26 | 102 | GGGAUAC<u>AAAA</u>GUAUCCA | 78.9 | 80.2 | 1.3 |
| RH27 | 102 | GGGAUAC<u>AAAAA</u>GUAUCCA | 79.1 | 79.0 | 0.1 |
| RH28 | 102 | GGGAUAC<u>AAAAAAA</u>GUAUCCA | 78.1 | 76.2 | 1.9 |
| RH29 | 102 | GGGAUAC<u>AAAAAAAAA</u>GUAUCCA | 73.3 | 72.2 | 1.1 |
| RH30 | 105,106 | GGCGCGG<u>CA CCGU</u>CCGCG GAACAAACGG [b] | 96.0 [c] | 98.0 | 2.0 |

[a] The sequences of hairpins are shown and the loop sequences are underlined. [b] The RNA is the hairpin that partially denatured from the pseudoknot of beet western yellow virus and the stability of the hairpin is studied in this work. [c] $T_m$ is taken from Fig. 7 in Ref. 106.